\documentclass[reprint, amsmath,amssymb,aps]
{revtex4-1}
\usepackage{graphicx}
\usepackage{dcolumn}
\usepackage{bm}
\usepackage{float}
\begin{document}

\title{All-Optical Spin Locking in Alkali-Vapor Magnetometers}

\author{Guzhi Bao$^{1,2}$}
\author{Dimitra Kanta$^{2}$}
\author{Dionysios Antypas$^{3}$}
\author{Simon Rochester$^{4}$}
\author{Kasper Jensen$^{5}$}
\author{Weiping Zhang$^{6,7}$}
\author{Arne Wickenbrock$^{2}$} 
\author{Dmitry Budker$^{2,3,8,9}$}

\affiliation{
$^{1}$Department of Physics, East China Normal University, Shanghai 200062, China. \\ 
$^{2}$Johannes Gutenberg-Universit{\"a}t  Mainz, 55128 Mainz, Germany\\
$^{3}$Helmholtz Institut Mainz, 55099 Mainz, Germany\\
$^{4}$Rochester Scientific, LLC., El Cerrito, CA 94530, USA\\
$^{5}$School of Physics and Astronomy, University of Nottingham, University Park, Nottingham NG7 2RD, England, United Kingdom\\
$^{6}$School of Physics and Astronomy, Shanghai Jiao Tong University, 
and Tsung-Dao Lee Institute, Shanghai 200240, China \\
$^{7}$Collaborative Innovation Center of Extreme Optics, Shanxi University, Taiyuan, Shanxi 030006, China\\
$^{8}$Department of Physics, University of California, Berkeley, CA 94720-7300, USA\\
$^{9}$Nuclear Science Division, Lawrence Berkeley National Laboratory, Berkeley, CA 94720, USA}
\date{\today}

\begin{abstract}
The nonlinear Zeeman effect can induce splittings and asymmetries of magnetic-resonance lines in the geophysical magnetic-field range. We demonstrate a scheme to suppress the nonlinear Zeeman effect all optically based on spin locking.
Spin locking is achieved with an effective oscillating magnetic field provided by the AC Stark-shift of an intensity-modulated and polarization-modulated laser beam. This results in the collapse of the multi-component asymmetric magnetic-resonance line with $\sim\,100\,$Hz width in the Earth-field range into a peak with a central component width of 25\,Hz. The technique is expected to be broadly applicable in practical magnetometry, potentially boosting the sensitivity and accuracy of Earth-surveying magnetometers by increasing the magnetic-resonance amplitude and decreasing its width. Advantage of an all-optical approach is the absence of cross-talk between nearby sensors when these are used in a gradiometric or in an array arrangement. 

\end{abstract}
\maketitle
\section{Introduction}

Measurements of magnetic fields with femtotesla sensitivity are critical to many applications including  geophysics \cite{H.B.Dang(2010)}, fundamental physics \cite{G.Vasilakis(2009),I.Altarev(2009)} and medicine \cite{G.Bison(2009),C.N.Johnson(2013)}. Optical magnetometers currently reach subfemtotesla$/\sqrt{\textrm{Hz}}$ sensitivity levels for submicrotesla fields \cite{H.B.Dang(2010),M.P.Ledbetter(2008),W.C.Griffith(2010),S.J.Smullin(2009),D.Budker(2007)}. However, in the geophysical field range (up to 100$\,\mu$T), the main limitation to the magnetic-resonance linewidth and sensitivity is the nonlinear Zeeman (NLZ) splitting \cite{S.J.Seltzer(2007),V.Acosta(2006),K.Jensen(2009),li2016unshielded}. The NLZ effect can cause splitting of resonance components, leading to a magnetometer signal decrease and to a spurious dependence of scalar-sensor readings on the relative orientation of sensor and magnetic field.

NLZ shifts can be effectively canceled with the use of double-modulated synchronous optical pumping \cite{S.J.Seltzer(2007)}, high-order polarization moments \cite{V.Acosta(2006)}, and tensor light-shift effects \cite{K.Jensen(2009)}. Recently, a new scheme to suppress the NLZ effect by adding a so-called spin-locking field \cite{Guzhi.B(2017)} was demonstrated. An oscillating magnetic field (RF field) or short magnetic-field pulses in the laboratory frame result, for appropriate parameters, in an effective magnetic field along the atomic magnetization in the rotating frame. Spins that are not pointing along this magnetic field precess around it, i.e. they are “locked”. As a result, the spin-locking field prevents splitting, shifts and lineshape asymmetries from occurring.  However, globally applied  magnetic fields may lead to crosstalk  between closely located sensors (as in a gradiometer) and therefore limit the applicability of this technique to sensor networks, which are important in biomedical and fundamental physics applications, such as, human heart or brain-activity mapping \cite{Bison.G(2009),Lembke.G(2014),Groeger.S(2005),Aleksandrov.EB(2006),Borna.A(2017)}. 
Additionally, in remote magnetometry applications, ``real" spin-locking magnetic fields cannot be directly applied to the atomic sample \cite{Higbie.M(2011),Patton.B(2012),bustos2018}. However, it is possible to apply fictitious magnetic fields.

In the presence of light, the energies of Zeeman sublevels are subject to  ``AC Stark-shifts" or ``light-shifts" \cite{Mathur(1968),Cohen-Tannoudji(1972),LeKien(2013),patton2014}. There are, depending on the polarization of the light and the atomic transition, scalar, vector and tensor shifts. In particular, the effect of the vector light shift (VLS) is analogous to a fictitious magnetic field \cite{Cohen-Tannoudji(1972),Zhivun(2016)}. VLS were studied in the context of all-optical magnetometry \cite{D.Budker(2007),opticallypolarizedatoms,patton2014}; in particular, light was used to substitute for RF fields \cite{zhivun(2014),lin(2017)}. 
In this paper, we present all-optical compensation of nonlinear Zeeman shift in a magnetometer using spin locking by replacing the RF field with an intensity- and polarization-modulated laser beam. This method allows to build a highly-sensitive multi-sensor magnetometer array, capable of working in the Earth's magnetic field range.

\section{Theoretical description}
\subsection{Nonlinear Zeeman effect}
The ground-state Hamiltonian of an atom in a magnetic field for states with electronic angular momentum $J = 1/2$ including both the hyperfine and Zeeman interactions, is:
\begin{equation}
\hat{H}=A_{J}\boldsymbol{I\cdot J}+g_{s}\mu_{B}\boldsymbol{S\cdot B}-g_{I} \mu_{N}\boldsymbol{I\cdot B},
\label{Eq:1}
\end{equation}
where $A_{J}$ is the hyperfine coupling constant, $g_{s}$ and $g_{I}$  are respectively the electron-spin and nuclear Land\'e factors
of the atom, $I$ is the nuclear spin,  $\mu_{B}$ is the Bohr magneton, and $\mu_{N}$ is the nuclear magneton. The first term describes the hyperfine interaction and the second and third terms describe Zeeman interactions. For a system with one valence electron in an $S (J=1/2)$ level,
the analytical solution for the eigenvalues of  the Hamiltonian yields the Breit-Rabi formula, which provides the energy shifts of the magnetic sublevels $\vert m\rangle$ for a state with a total angular momentum $F$ in a magnetic field of strength $B$ \cite{G.Breit(1931),opticallypolarizedatoms}:
\begin{equation}
E_{m}=\frac{\Delta_{hf}}{2(2I+1)}-g_{I}\mu_{B}mB\pm\frac{\Delta_{hf}}{2}(1+\frac{4m\xi}{2I+1}+\xi^{2})^{1/2},
\label{Eq:Breit-Rabi}
\end{equation}
where $\xi=(g_{J}+g_{I})\mu_{B}B/\Delta_{hf}$, $\Delta_{hf}$ is the hyperfine-structure interval and the $\pm$ sign refers to the $F = I\pm1/2$ hyperfine levels. Relevant to our studies, is that the nonlinear term in Eq.\,\eqref{Eq:Breit-Rabi} is important under earth field. We expand the eigenenergies as a series in $B$ around 0. The transition frequencies corresponding to $\Delta\,m = 1$ for the cesium \,$6^{2} S_{1/2}\,F=4$ system are:
\begin{equation}
E_{m+1}-E_{m}\approx\frac{\mu_{B}B}{4}+ \frac{(\mu_{B}B)^{2}}{16\Delta_{hf}}(2m-1),
\end{equation}
where $(\mu_{B}B)^{2}/16\Delta_{hf}$ is the so-called quantum-beat revival frequency \cite{S.J.Seltzer(2007)}, $g_{J}\approx\,2$ and we neglect the Zeeman energy of the nuclear spin [last term in Eq.\,\eqref{Eq:1}] as $\mu_N << \mu_B$. Assuming the system is working in Earth's magnetic field which is around 50$\,\mu$T, the calculated revival frequency is $\omega_{rev}=2\pi\cdot3.3\,$Hz 
\cite{opticallypolarizedatoms}. 
This frequency is comparable to the magnetic resonance width and hence the system is strongly affected by nonlinear Zeeman effect. The Cs magnetic resonance is split into eight peaks for a magnetic field in the Earth-field range (see Fig.\,\ref{fig:intensitymodulation}). This mechanism broadens the linewidth while reducing the signal amplitude and, consequently, reduces the magnetometer sensitivity.

\subsection{Spin locking}
To describe the physics of spin-locking, we start with a total angular momentum $F=1$ system interacting with a leading magnetic field along $\hat{z}$ and an oscillating magnetic field along $\hat{x}$. 
We further assume that the atomic spins are initially prepared in the $m_F=1$ state along the $\hat{x}$ direction by a circularly polarized pump field and that the probe-light power is sufficiently low to be neglected for the dynamics.

\begin{figure}[tbph]
\centering
\includegraphics[width=8.6cm]{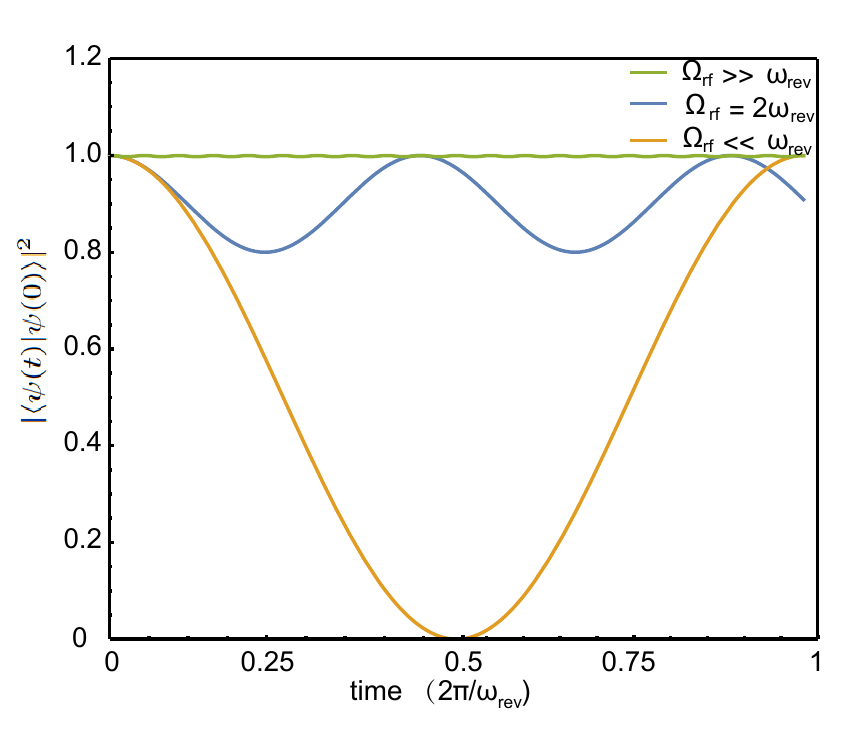}
\caption{Probability $P(t,0)$ for an atom to be found in the initial state. For small amplitudes of the spin-locking field ($\Omega_{rf}\ll\,\omega_{rev} $), the probability undergoes quantum beating. If the amplitude of the spin-locking field is much larger than the NLZ effect ($\Omega_{rf}\gg\,\omega_{rev}$), the atoms remain in the initial state. If the amplitude of the spin-locking field is equal to the NLZ parameter ($\Omega_{rf}=2\omega_{rev}$), the atoms are partially locked in the initial state and the populations undergo oscillation with frequency $\omega_{lock}$.}  
\label{fig:4NLZ}
\end{figure}

The Hamiltonian of the system in the basis of the Zeeman sublevels with the quantization axis along $\hat{z}$ is:
\begin{equation}
\hat{H}_{SL}=\hbar\begin{pmatrix}
\Omega_{L}+\omega_{rev}&-\frac{\Omega_{rf}\sin\left(\omega_{rf}t+\phi\right)}{\sqrt{2}}&0\\
-\frac{\Omega_{rf}\sin(\omega_{rf}t+\phi)}{\sqrt{2}}&0&-\frac{\Omega_{rf}\sin(\omega_{rf}t+\phi)}{\sqrt{2}}\\
0&-\frac{\Omega_{rf}\sin(\omega_{rf}t+\phi)}{\sqrt{2}}&-\Omega_{L}+\omega_{rev}\\
\end{pmatrix},\\
\end{equation}
where $\Omega_{L}$ is the Larmor frequency proportional to the leading field; $\omega_{rev}$ characterizes the strength of the NLZ effect; $\Omega_{rf}$ is the Larmor frequency corresponding to the oscillating field and proportional to its amplitude; $\omega_{rf}$ and $\phi$ are the frequency and phase of the oscillating field, respectively. We perform a transformation into the frame rotating at $\omega_{rf}$ by means of the unitary operator $U(t)\,=\,\textrm{exp}\,(−iH^{'}t)$, where
\begin{equation}
H^{'}=
\begin{pmatrix}
\omega_{rf}&0&0\\
0&0&0\\
0&0&-\omega_{rf}
\end{pmatrix}.\\
\end{equation}
Applying the rotating-wave approximation removes the time dependence from the Hamiltonian; the oscillating field appears as a static magnetic field, whose direction depends on the phase $\phi$. To understand spin-locking, we assume a magnetic field along the precessing spins ($\omega_{rf}=\Omega_{L}$, $\phi=\pi/2$). The Hamiltonian in the rotating frame is then:
\begin{equation}
\hat{H}_{rotate}=\hbar\begin{pmatrix}
\omega_{rev}&-\frac{\Omega_{rf}}{2\sqrt{2}}&0\\
-\frac{\Omega_{rf}}{2\sqrt{2}}&0&-\frac{\Omega_{rf}}{2\sqrt{2}}\\
0&-\frac{\Omega_{rf}}{2\sqrt{2}}&\omega_{rev}
\end{pmatrix}.\\
\end{equation}
We write the state $\vert \psi(t)\rangle$ of $F$ as a superposition of energy eigenstates $\Psi_{i}$ with eigenvalues $E_i$:
\begin{equation}
\vert \psi(t)\rangle=\sum_{i}\Psi_{i} e^{\frac{-i}{\hbar}E_{i}t}.
\end{equation}
The probability $P(t,0)$ for an atom to be found in the initial state, $\vert\langle\psi(t)\vert\psi(0)\rangle\vert^{2}$, can be written as:
\begin{equation}
P(t,0)=\frac{\omega_{lock}^{2}+\Omega_{rf}^{2}+\omega_{rev}^{2}\cos(\omega_{lock}t)}{2\omega_{lock}^{2}},
\end{equation}
where $\omega_{lock}=\sqrt{\omega_{rev}^{2}+\Omega_{rf}^{2}}$ is the oscillating frequency of $P(t,0)$.
With an increase of the spin-locking field amplitude, the oscillating component of $P(t,0)$ decreases in amplitude and the frequency of the oscillation increases (see Fig.\,\ref{fig:4NLZ}). In this simplified model, it appears that spin locking improves with the amplitude of the applied field. However, we observe, that the presence of the locking field leads to power broadening of the magnetic resonance; this results in an optimal amplitude of the field such that $\Omega_{rf}$ is comparable to $\omega_{rev}$.
\subsection{Spin Locking with AC Stark shift}
Vector light shift gives rise to a fictitious magnetic field along the light propagation direction with amplitude \cite{julsgaard2003entanglement}:
\begin{equation}
B_{\textrm{fict}}=\frac{-c(\Delta) I \epsilon }{\hbar \gamma},
\end{equation}
where $c(\Delta)$ is a proportionality constant which depends on atomic parameters and the frequency detuning $\Delta$ from atomic resonance, $I$ is the light intensity, $\gamma$ is the gyromagnetic ratio and $\epsilon=\left[ I(\sigma_+)-I(\sigma_-) \right]/I$ is the  Stokes parameter specifying the degree of circular light polarization ($\epsilon=+1$ for $\sigma_+$-polarized light, $\epsilon=-1$ for $\sigma_-$-polarized light and $\epsilon=0$ for linear-polarized light).

In an $F=1\,$to$\,F'=0$ transition system, we induce a circularly polarized light-shift field propagating along the pump ($\hat{x}$ direction). The light-atom interaction Hamiltonian under the rotating-frame approximation is:
\begin{equation}
\hat{H}_{I}=\hbar\begin{pmatrix}
0&0&0&\frac{i\Omega_{LS}}{2\sqrt{3}}\\
0&0&0&-\frac{\Omega_{LS}}{\sqrt{6}}\\
0&0&0&\frac{i\Omega_{LS}}{2\sqrt{3}}\\
-\frac{i\Omega_{LS}}{2\sqrt{3}}&-\frac{\Omega_{LS}}{\sqrt{6}}&-\frac{i\Omega_{LS}}{2\sqrt{3}}&-\Delta
\end{pmatrix}.\\
\end{equation}
With this perturbation, the corrections of ground state Hamiltonian is:
\begin{equation}
\hat{H}_{I}^{'}=\frac{\Omega_{LS}^{2}}{\Delta}\begin{pmatrix}
-\frac{1}{48}&-\frac{1}{24\sqrt{2}}&\frac{1}{48}\\
\frac{1}{24\sqrt{2}}&-\frac{1}{24}&\frac{1}{24\sqrt{2}}\\
\frac{1}{48}&\frac{1}{24\sqrt{2}}&-\frac{1}{48}\\
\end{pmatrix}.\\
\end{equation}
The light-shift beam can be intensity- and/or polarization-modulated (see the details in the experimental section below). For the intensity-modulation scheme, the total Hamiltonian for the ground-state evolution is:
\begin{equation}
\hat{H}_{tot}=\hat{H}_{SL}+\lbrack\,cos(\omega_{rf}\,t)+1\rbrack\,\hat{H}_{I}^{'}.
\end{equation}
To compare the optical rotation signal with rf field or light-shift field (see below), we keep the oscillating-magnetic-field terms in $\hat{H}_{SL}$.
\subsection{Optical Rotation Signal}
\begin{figure}[tbph]
\centering
\includegraphics[width=8.6cm]{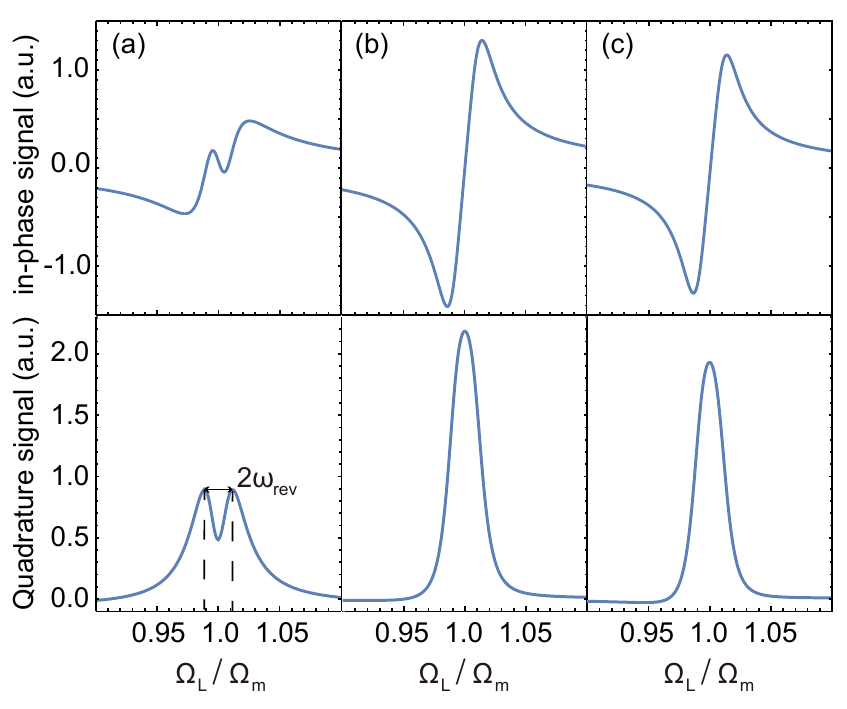}
\caption{Theoretical calculated in-phase (top row) and quadrature (bottom row) first-harmonic amplitudes of optical rotation signal. With NLZ effect, the magnetic resonance is split into two peaks (a); with rf spin locking field (b) or intensity modulated light-shift field (c), the spin is locked and magnetic resonance has only one peak. For these plots, the parameters $\Omega_{rf}/\omega_{rev}\,=30$, $\Omega_{LS}/\omega_{rev}\,=800$ are chosen.}  
\label{fig:theory}
\end{figure}
Let us assume that probe light linearly polarized along $\hat{x}$ with frequency $\omega$ is used to measure the atomic state of Cs during its evolution. The propagation direction $\hat{y}$ of probe is perpendicular to both the propagation direction of the pump light $\hat{x}$ and the direction of leading field $\hat{z}$. After probe light propagates through the medium with polarization $\bm{P}=n\textrm{Tr}\,\rho\,\bm{d}\,=\,\textrm{Re}\{e^{i(\hat{y}\cdot\,\hat{r}-\omega\,t+\phi)}[(P_{1}-iP_{2})\hat{x}+(P_{3}-iP_{4})\hat{z}]\}$ ($n$ is the atomic density, $\rho$ is the density matrix of atomic ensemble, $\bm{d}$ is dipole operator, $P_{i}$ are the in phase and quadrature components of the polarization), the light polarization $\alpha$ change as \cite{opticallypolarizedatoms}：
\begin{equation}
\frac{d\alpha}{dz}=\frac{2\pi\omega}{\varepsilon_{0}c}P_{4},
\end{equation}

where $\varepsilon_{0}$ is the electric field amplitude. The pump field is periodically 
modulated with frequency $\Omega_{m}$. To simplify the calculation, here we assume pump is sinusoidally modulated. When $\Omega_{m}\,=\omega_{rf}$, we can solve the time-periodic evolution equation by using Floquet theory \cite{rochester2010modeling,coop2017floquet}.
Figure\,\ref{fig:theory} shows in-phase and quadrature first-harmonic amplitudes of the optical-rotation signal. Without the light-shift beam (a), the magnetic resonance is split due to the NLZ effect. Figures\,\ref{fig:theory}\,(b) and (c) shows the magnetic resonance with rf field and amplitude-modulated light-shift field, respectively. Spin locking is achieved with an RF-magnetic field \cite{Guzhi.B(2017)} is seen in Fig.\,\ref{fig:theory}\,(b). In Fig.\,\ref{fig:theory}\,(c), we consider $\sigma_+$-polarized light ($\epsilon = +1$) with its intensity fully modulated at frequency $\Omega_{LS}$. In this case, the light provides a fictitious magnetic field $B_{\rm{fict,B}} \propto\,I_0\,[\cos\,\left(\,\omega_{rf} t\,\right)+1]$.
\section{experimental apparatus}
\begin{figure}[tbph]
\centering
\includegraphics[width=8.6cm]{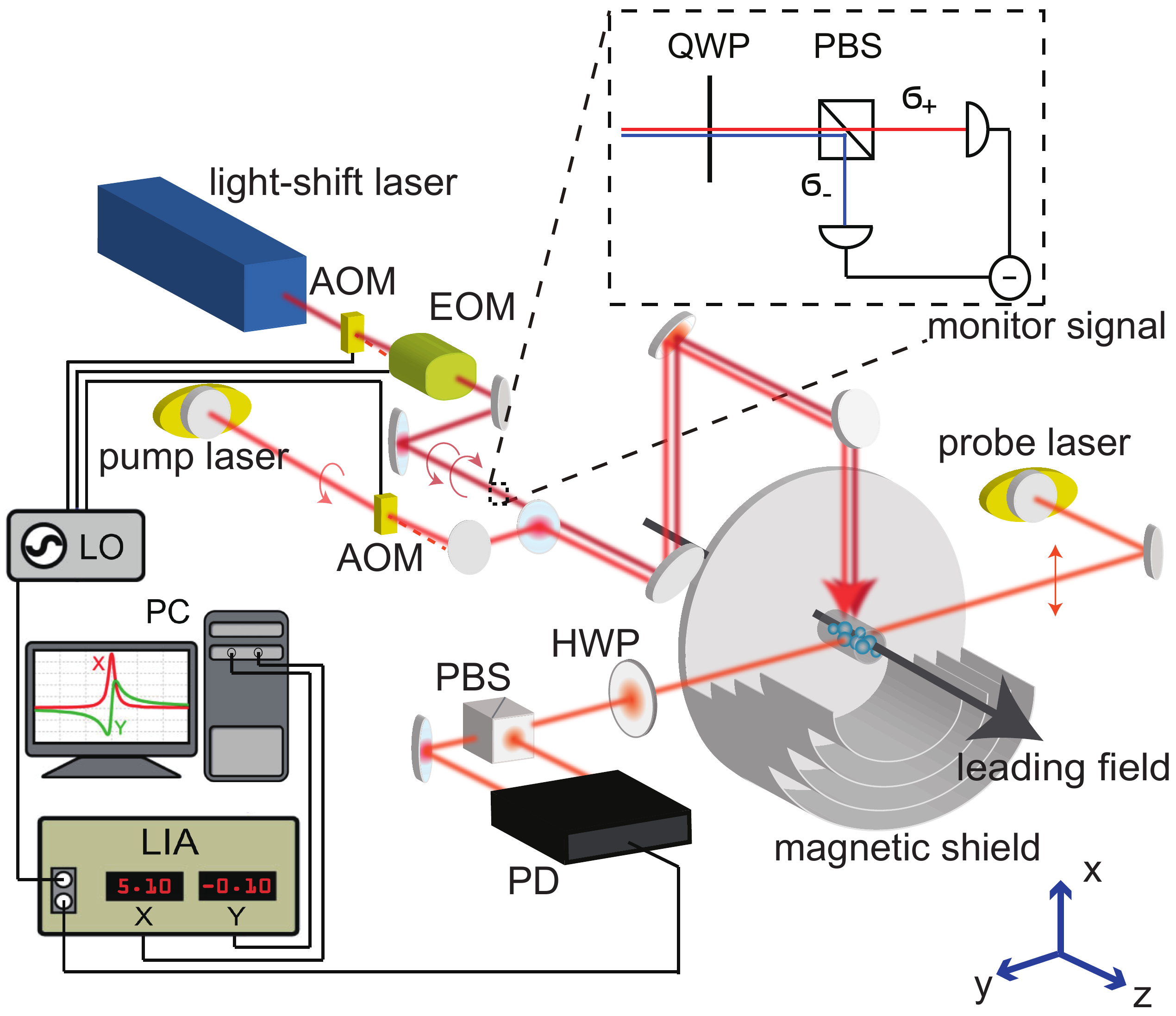}
\caption{Experimental setup. AOM: acousto-optic modulator; EOM: electro-optic modulator; HWP: half-wave plate; PBS: polarizing beamsplitter; PD: balanced photodetector; LIA: lock-in amplifier; LO: local oscillator. A partial view of the magnetic shield is shown in the figure. Atoms are contained in a glass cell positioned in the center of the magnetic shield and are pumped (along $\hat{-x}$) and probed (along $\hat{y}$) by laser beams under a static magnetic field (along $\hat{z}$). The intensity of the light-shift laser beam is sinusoidally modulated with an AOM at a frequency $\Omega_{m}$, while its polarization is switched between the $\sigma_{+}$ and $\sigma_{-}$ states every $\pi/\Omega_{m}$, using an EOM. Inset shows the monitor setup for polarization and amplitude of the light-shift beam.}
\label{fig:apparatus}
\end{figure}
Figure\,\ref{fig:apparatus} shows the experimental apparatus. A paraffin-coated cylindrical cell \cite{H.G.Robinson(1958)cell,M.A.Bouchiat(1964)cell,E.B.Alexandrov(1992)cell,E.B.Alexandrov(2002)cell} with a length of 5\,cm and a diameter of 4\,cm  containing $^{133}$Cs at room temperature, is enclosed within a four-layer mu-metal magnetic shield. The atoms are prepared in the stretched state along the $-\hat{x}$ axis by optical pumping with a circularly polarized, $-\hat{x}$-directed laser beam. The pump-laser frequency is locked to the Cs\,D2\,$6^{2}S_{1/2}\ F = 3\rightarrow\  6^{2}P_{3/2}\ F^{'}  = 4$ transition at 852\,nm with a dichroic atomic vapor laser lock (DAVLL) \cite{V.V.Yashchuk(2000)laserlock}.
The intensity of this beam is pulsed (3\% duty cycle) with an acousto-optic modulator (AOM). The light power during the ``on'' part of the cycle is 50$\,\mu$W.
Polarization oscillations of a $10\,\mu W$, $\hat{y}$- directed probe beam induced by the polarized atomic vapor are measured with a balanced polarimeter upon transmission through the cell. The beam is linearly polarized along the $\hat{x}$-axis and detuned by about 4\,GHz towards higher frequencies of the Cs\,D2\,$F = 4\rightarrow  F^{'} = 5$ transition.
A circular-polarized light-shift beam produced with a Ti:sapphire laser propagates parallel to the pump beam. The intensity of the beam is modulated with an AOM and its polarization is modulated with an EOM in order to provide time-varying light shift. The frequency of this laser can be widely tuned and is, for most of the experiments presented here, detuned by 10\,GHz from the D2\,$6^{2}S_{1/2}\,F\,=\,4\rightarrow\,6^{2}P_{3/2}\,F^{'}\,=\,5$ transition towards lower frequencies. Its frequency is stabilized to the internal reference cavity of the laser. The detuning of 10\,GHz was chosen to minimize optical pumping by the light-shift beam while maintaining sufficient fictitious magnetic field amplitude ($\approx$14\,nT for 250\,mW power) for spin-locking.

To measure the magnetic resonance, we fix the modulation frequency $\Omega_{m}$ of both pump and light-shift beams at a particular value (corresponding to Larmor frequencies for magnetic fields of up to 100$\,\mu$T). We scan the leading $\hat{z}$-directed magnetic field and therefore the Larmor frequency and observe the polarization of the probe beam. The signal from the balanced polarimeter is fed into a lock-in amplifier and demodulated at the modulation frequency. 
The magnetic resonance can be observed in the polarization rotation amplitude and phase of the probe beam \cite{D.Budker(2002)}. 

The setup to monitor the modulation of the light-shift beam is shown in the inset of Fig.\,\ref{fig:apparatus}. A QWP is used to convert circular into linear polarization. The resulting $\sigma_{+}$ and $\sigma_{-}$ components are split via PBS and sent to two photo-detectors. 
\section{experimental results}
\begin{figure*}[tbph]
\centering
\includegraphics[width=17.2cm]{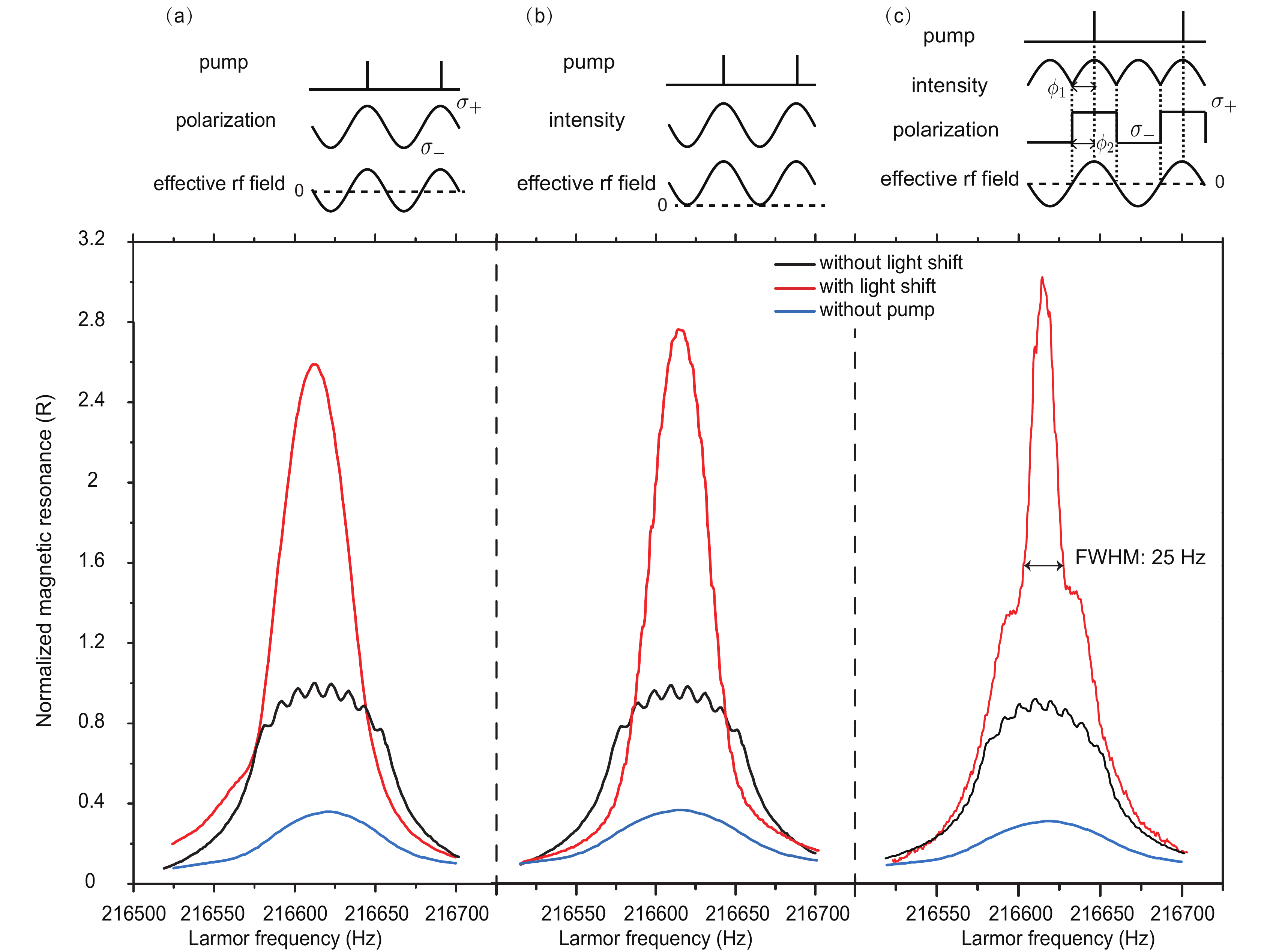}
\caption{Magnetic-resonance lineshape for a modulation frequency of 216,620\,Hz as a function of the leading magnetic field along the $\hat{z}$-axis with applied light-shift field and pump (red line), without light-shift field (black line), and without pump field (blue line). The amplitude of magnetic resonance without light shift is normalized to one. The power in (a) and maximum power in (b,c) of the light-shift beam is 200\,mW. The inset shows the polarization modulation (a), intensity modulation (b) and modulated both (c) scheme for the pump and light-shift field.}
\label{fig:intensitymodulation}
\end{figure*}
We deployed three different methods to modulate the light-shift beam and achieve spin locking. 
Figure\,\ref{fig:intensitymodulation} shows the amplitude of the lock-in output as a function of the leading magnetic field around 60\,$\mu$T with the pump-laser modulation frequency fixed at 216,620\,Hz. The magnetic resonance spectra are shown without and with application of the light-shift beam (black and red curves, respectively), as well as without the pump beam (blue curve). 

In the method depicted in Fig.\,\ref{fig:intensitymodulation} (a,b), either the intensity or the polarization of the light-shift beam is modulated, to provide a sine-modulated light shift, as in Refs.\cite{zhivun(2014),patton2014}. In the polarization-modulation scheme, the fictitious magnetic field is oscillating around zero. However, in this scheme, the light is only purely circularly polarized when $\epsilon=\pm\,1$; the presence of the other polarization states cause tensor-light shifts result broadening of the linewidth. In the amplitude-modulation scheme, the VLS produces a fictitious magnetic field of magnitude $B_{fict}\propto[1+\text{cos}(\Omega_{m}t)]$. 
The oscillating term of the fictitious field locks the spins. The static term of the fictitious field plays no role in spin-locking but the constant light which leads to, for example, broadening of the linewidth due to residual optical pumping. In the absence of the light-shift beam, the magnetic resonance is split into eight partially-resolved Lorentzian peaks, due to the NLZ effect. Applying the modulated light-shift beam results in a 
narrower full-width-half-maximum (FWHM) central peak and an 
amplitude increase.
\begin{figure}[tbph]
\centering
\includegraphics[width=8.6cm]{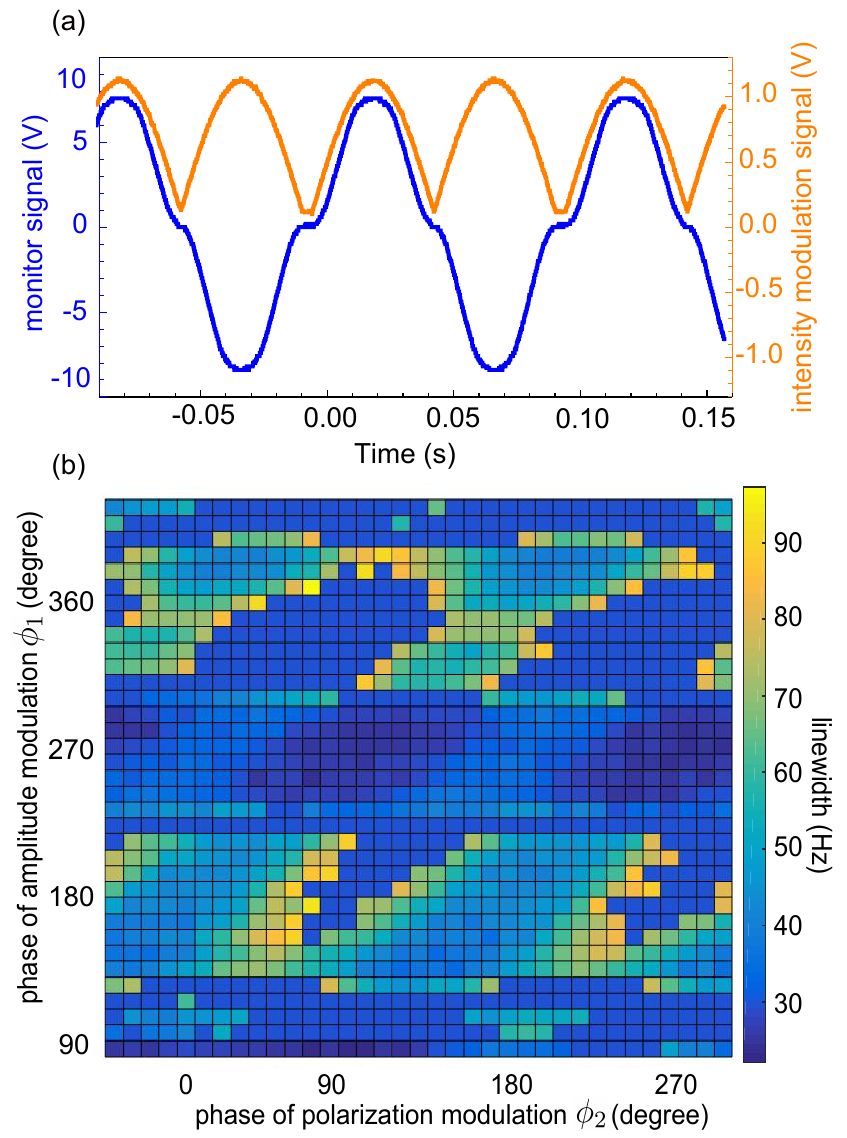}
\caption{(a) Observed monitor signal and input intensity modulation signal. Here $\phi_{1}=-\pi/2$ and $\phi_{2}=-\pi/2$. The distortion of fictitious RF field is mainly caused by AOM's nonlinear response. (b) Phase scanning of $\phi_{1}$ and $\phi_{2}$.}
\label{fig:modulationscheme}
\end{figure}

In Fig.\,\ref{fig:intensitymodulation} (c), the direction of the fictitious magnetic field is changed by modulating the polarization of the light-shift beam from $\sigma_{+}$ to $\sigma_{-}$ using an EOM. The intensity\,$I$ of the light-shift beam is modulated with an AOM as $I\,\propto\,\textrm{Abs}[\cos(\Omega_{m}t)$]. In this modulation scheme, neglecting the counter-rotating component, the fictitious RF field actually co-rotates with the precessing spins. 

The phase between intensity and polarization modulation of the VLS beam needs to be chosen carefully to create a pure fictitious RF field. Additionally, to enable spin-locking, the fictitious RF field has to be in-phase with the pump pulse to ensure that it points along the direction of the precessing spins. We show the monitor signal (produced by subtraction of the $\sigma_{+}$ and $\sigma_{-}$ recorded powers) for $\phi_{1}=-\pi/2$ and  $\phi_{2}=-\pi/2$ in Fig.\,\ref{fig:modulationscheme}\,(a), as an fictitious RF field. Here $\phi_{1}$ is the phase of the intensity modulation and $\phi_{2}$ is the phase of the polarization modulation. Figure\,\ref{fig:modulationscheme}\,(b) displays the magnetic-resonance linewidth for different $\phi_{1}$ and $\phi_{2}$. The best results are achieved around (combinations of) $\phi_{1}=\pi/2,\,3\pi/2$ and $\phi_{2}=\pi/2,\,3\pi/2$. 

\begin{figure}[tbph]
\centering
\includegraphics[width=8.6cm]{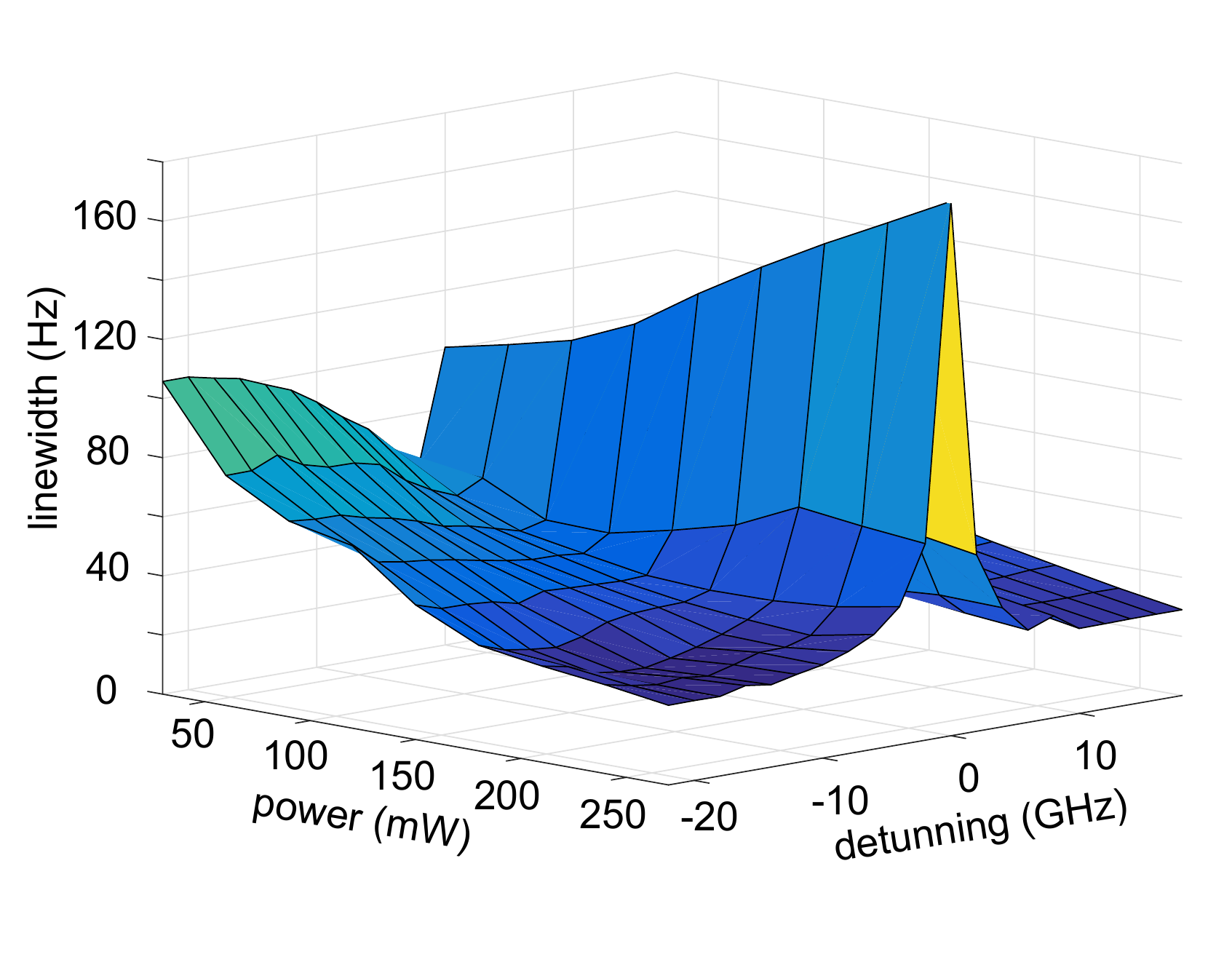}
\caption{Map of the magnetic resonance linewidth as a function of the applied light-shift-field power and detuning. The minimum linewidth is 25.25\,(6)\,Hz.}
\label{fig:poweranddetunning}
\end{figure}

Figure\,\ref{fig:poweranddetunning} shows the effective linewidth of the magnetic resonance at earth field (60\,$\mu$T) as a function of the applied light-shift beam power and detuning. When the light-shift beam is of low power and detuned far off-resonance, there is no spin locking and the effective linewidth is $\approx\,100$\,Hz.  
When the light-shift beam is near resonance with the atomic transition, the effect of optical pumping is much stronger than that of the VLS. As a result, the linewidth of the magnetic resonance is even broader than that observed in the absence of the light-shift beam. When the light is far-off resonant from the optical transition, the optical pumping is negligible and the VLS dominates the interaction. We observe a minimum of the linewidth for a 220\,mW light-shift beam, 10\,GHz detuned below the D2\,$6^{2}S_{1/2}\,F\,=\,4\rightarrow\,6^{2}P_{3/2}\,F^{'}\,=\,5$ transition. Note, however, that spin locking works well also for the opposite sign of detuning, corresponding to a sign reversal of the effective RF field. The power applied was limited by the available laser.

\section{Conclusion}
An all-optical method to suppress the NLZ effect in the range of the Earth's magnetic field using spin locking is demonstrated.
A polarization and intensity modulated light-shift beam is applied which effectively suppresses NLZ-related broadening of the magnetic resonance. The effect also works with individual application of intensity or polarization modulation but the combination of both yields the best result. In contrast to other techniques, this method does not cause any crosstalk in sensor networks and also does not interact with samples close to the sensor. The Larmor frequency of the optimal effective spin-locking field in the rotating frame is comparable to the spin-revival frequency; the phases ($\phi_{1}$ and $\phi_{2}$) are chosen such that the co-rotating part of the fictitious RF magnetic field is colinear with the precessing spins. We note, that with the sensitivity of Earth-field magnetometers is improved for two reasons: due to the increase in the magnetic-resonance signal amplitude and due to the reduction in the effective linewidth. The area of the magnetic resonance profile with both pump and light shift beam is larger than the sum of the profile areas corresponding with only pump or light shift beam. This effect might be arising from pumping and repumping by the light shift beam and needs to be further studied. In the current setup, we observe some linewidth broadening due to optical pumping by the light-shift beam. The efficiency of the all-optical spin-locking scheme can be further improved using higher power and increased optical detuning.  

The authors acknowledge useful discussions with Nataniel L. Figueroa and financial support by the German Federal Ministry of Education and Research (BMBF) within the “Quantumtechnologien” program (FKZ 13N14439) and the DFG through the DIP program (FO 703/2-1). Guzhi Bao acknowledges support by the China Scholarship Council. Weiping Zhang acknowledges support from National Key Research and Development Program of China under Grant No. 2016YFA0302001, the National Natural Science Foundation of China (Grants No. 11654005, No. 11234003, and No. 11129402), and the Science and Technology Commission of Shanghai Municipality (Grant No. 16DZ2260200). 

\bibliography{ref.bib}
\end{document}